\documentclass[pre,aps,showpacs,twocolumn]{revtex4}
\usepackage{amsmath,amsfonts,bm}
\usepackage[dvips]{graphicx}

\begin{document}

\title{Polymer dynamics in chaotic flows with strong shear component}
\author{K. S. Turitsyn}
\affiliation{Landau institute for Theoretical Physics, Moscow,
Kosygina 2, 119334, Russia\\
Theoretical Division, LANL, Los Alamos, NM 87545, USA}
\date{\today}
\begin{abstract}
We consider the internal dynamics of the polymer molecule which is
injected in the chaotic flow with strong mean shear component. The
flow geometry corresponds to the recent experiments on the elastic
turbulence (Groisman, Steinberg 2000). The passive polymer in such
flows experiences aperiodic tumbling. We present a detailed study
of the statistical properties of such polymer dynamics. First we
obtain the stationary probability distribution function of the
polymer orientation. Secondly we find the distribution of the time
periods between consequent events of tumbling, and finally we find
the tails of the polymer size distribution.
\end{abstract}
\pacs{83.80.Rs,47.27.Nz}

\maketitle Hydrodynamics and rheology of dilute polymer solutions
has attracted much theoretical and experimental attention
recently. Addition of small amount of polymers to ordinary liquid
leads to crucial changes of liquid properties. One of the most
famous effects of this type is the phenomenon of drag reduction.
The addition of few parts per million of long-chain polymer
molecules produces a dramatic reduction of the friction drag.
Although this effect was first observed by Toms in 1949
\cite{Toms}, there is still no rigorous theory, explaining the
phenomena. The qualitative description was proposed by Lumley
\cite{Lumley1,Lumley2}, but no quantitative theory is available.
Another spectacular phenomena, observed in dilute polymer
solutions is the effect of elastic turbulence, discovered recently
by Groisman and Steinberg \cite{00GS,04GS}. In this experiment a
chaotic fluid motion was observed in the system with small
Reynolds number ${\rm Re} \ll 1$. Obviously, such behavior can not
be observed in Newtonian liquids, where the flow should be
laminar, so the chaotic flow is generated by the elastic
instabilities of polymer solution. The dynamics of polymers and
possible mechanisms, explaining the chaotic state were studied in
the recent theoretical works \cite{00BFL,00Che,03FL}. It was
proposed that elastic instabilities occur because of elongation of
a single polymer in external flows.  The analysis of the single
polymer dynamics in chaotic flows, shows that the transition
between two qualitatively different types of behavior of polymers
can be observed in such system. This transition is called
coil-stretch transition. It separates the dynamics in weak flows,
where polymer molecules remain in coiled state most of the time,
and strong flows, where the molecules become substantially
elongated. The measure of the flow strength is given by the
Weissenberg number which is the product of the characteristic
velocity gradient and the polymer relaxation time. More precisely
the Weissenberg number is defined as ${\rm Wi} = \lambda \tau$,
where $\lambda$ is the largest Lyapunov exponent, associated with
the flow, and $\tau$ is the relaxation time of the slowest polymer
excitation mode. The coil-stretch transition occurs at ${\rm Wi} =
1$. With the development of novel optical methods a number of high
quality experimental observations focusing on resolving dynamics
of individual polymers (DNA molecules) subjected to a
non-homogeneous flow have been reported
\cite{97PSC,98SC,99SBC,01HSBSC}. It made possible the direct
observation of coil-stretch transition \cite{04GCS}. However in
all theoretical works, describing the polymer dynamics in chaotic
flows one of the important assumptions was the isotropic
statistics of the velocity field. At the same time in the
experiments on elastic turbulence the flow consists of regular
(shear-like) and chaotic components, the latter is relatively weak
in comparison with the former one. Thus, it became an important
task to generalize the known theoretical results on the flows of
such type.

In this paper we study the dynamics of the DNA polymers injected
in the chaotic flow. The flow consists of large stationary shear
component with the small chaotic part. It is assumed that the
concentration of the injected polymers is small enough so that
their back reaction on the flow is negligible. It was shown in
\cite{04CKLT} that the polymer spends most of the time stretched
in the shear direction and experiences aperiodic tumbling towards
the opposite direction. The main aim of this paper is to present a
detailed investigation of the statistical properties of such
polymer dynamics and to make some predictions which can be checked
experimentally. First, we obtain the exact expression of the
polymer direction distribution. We show that the body of the
angular distribution function is located in the region of small
angles, which correspond to the polymer stretched in the shear
direction. However, the tails of the PDF are algebraic, so the
fluctuations of the polymer direction are very strong. Secondly we
study the statistics of the time periods between consequent events
of polymer tumbling. It is shown that this PDF has an exponential
tail at large tumbling times, and two different asymptotic regimes
in the region of very small times. Finally, for polymers, which
are below the coil-stretch transition we obtain the asymptotic of
the polymer size distribution function, which is also algebraic.
We also show, that mean-shear component leads to a significant
broadening of the polymer size distribution. This effect is
surprising at the first sight, because the shear component itself
doesn't lead to an exponential polymer growth, and thus can not
lead to the algebraic tails of the polymer size distribution.
However the combined effect of the strong regular shear and
chaotic velocity components leads to a significant enhancement of
the polymer growth rate.

The overall plan of this article is following: in first section we
describe the experimental setup in which the elastic turbulence is
observed, then we discuss the assumptions of the models,
describing internal polymer dynamics and flow statistics. In the
main sections we give the detailed analysis of the polymer
dynamics, including the tumbling phenomenon \cite{04CKLT} and
statistical properties of polymer elongation. While in this paper
we present only analytical predictions, some of them have been
successfully checked numerically in other paper \cite{alb}.

\section{Experimental setup}
The classical effect of well-developed turbulence is observed in the
systems with large values of Reynolds number ${\rm Re} = V L/\nu \gg
1$, where $V$ is the characteristic value of the flow velocity, $L$
is the characteristic system size, and $\nu$ is kinematic viscosity.
In this case the chaotic motion is generated by the large nonlinear
terms in the Navier-Stokes equation \cite{frisch}. In the recently
discovered elastic turbulence phenomena the chaotic behaviour is
observed at extremely small Reynolds numbers ${\rm Re}\ll 1$, and is
due to elastic instabilities of the polymer solution. The
experimental setup in which this effect has been observed looks as
following: a swirling flow is generated between two coaxial disks of
the radius $R$ and with the gap $d\ll R$. The upper disk is being
rotated with the angular frequency $\Omega$, while the lower one is
stationary. For small frequencies or usual Newtonian liquid the
following laminar stationary flow will be generated:
\begin{equation}\label{regular}
 V_\phi = \Omega r z/d,\qquad V_r=V_z=0,
\end{equation}
Here $\phi,z,r$ are the usual cylindrical coordinates with the
center coinciding with the center of the lower disk. In the dilute
polymer solution a small chaotic component of the velocity field $v$
appears after some critical value of $\Omega$ (but still at small
Reynolds numbers). Currently existing theoretical works are unable
to describe the statistical properties of such flow. Thus, it is of
interest to study this properties experimentally. One of the
experimental possibilities of such study is the observation of
dynamics of single long polymer molecule (such as DNA). It should be
stressed, that the observed molecules are not related to the
polymers, which are dissolved in the liquid, and which generate the
chaotic flow. The concentration of the observed DNA molecules is
negligibly small, so they can be treated as passive objects. In the
following text we won't analyze the origins of the elastic
turbulence, and will denote by the polymers the passive observed DNA
molecules.

\section{Polymer and chaotic flow models}
\begin{figure}[tl]
 \includegraphics[width=0.42\textwidth]{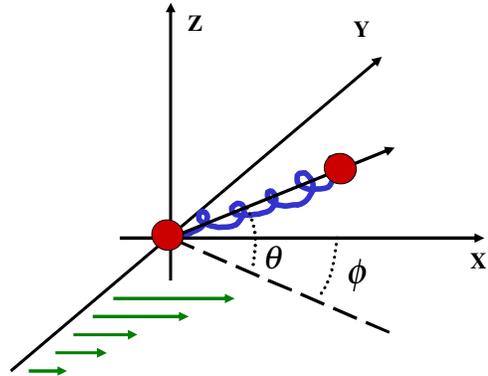}
 \caption{Schematic figure explaining polymer orientation geometry.}
 \label{fig:orient}
\end{figure}

For typical polymers the inertial effects can be neglected, and one can assume
that the polymer simply follows the lagrangian trajectories of the velocity field.
In this case the velocity field affects the internal polymer dynamics in two ways:
firstly the local velocity gradients lead to the polymer stretching and rotation,
secondly, if the flow parameters are spatially inhomogeneous (which happens for
example in the experimental setup described above), the lagrangian advection can
lead to the time dependent local velocity gradients, corresponding to the
stationary velocity component. In our analysis we neglect the second effect.
In the case of the elastic turbulence experimental setup described above, the
advection in the radial direction, which leads to the variations of the local
shear rate is determined only by the chaotic component which is assumed to be small.
It means the that significant changes of the local shear rate, due to lagrangian
advection occur on the time scales, much larger than the characteristic time scales
associated with the internal polymer dynamics. Therefore in the adiabatic approximation
one can assume the local shear rates, correposding to the regular velocity component
to be constant.
Polymer extension can be described by the simple dumb-bell model, where the end-to-end
separation vector {\bf R} satisfies the following equation \cite{Hinch,book}:
\begin{equation} \label{main}
 \partial_t R_i = R_j \nabla_j v_i - \gamma(R) R_i + \zeta_i
\end{equation}
where the relaxation rate $\gamma$ is the function of the absolute
value $R$ of the vector ${\bf R}$, the velocity gradient $\nabla_j
v_i$ is taken at the molecule position, and $\zeta_i$ is the thermal
Langevin force. We assume that velocity field is large-scale, so it
can be assumed to be smooth on the polymer size. This assumption
justifies, the linear approximation of the velocity field used in
(\ref{main}). In this paper we discuss two different situations.
Firstly when the Lyapunov exponent, corresponding to the velocity
field is larger than the polymer relaxation rate, so the
coil-stretch transition has already occurred, the polymer becomes
nonlinearly stretched so, one can neglect the thermal forces
$\zeta_i$ acting on him \cite{00BFL}. One can introduce the polymer
direction vector $n_i = R_i/R$, which is in this case described by
the following equation:
\begin{equation} \label{direct}
 \partial_t n_i  = n_j (\delta_{ik}-n_i n_k)\nabla_i v_k
\end{equation}
One can see that the direction evolution is completely decoupled from the
dynamics of the polymer size $R$. In the second situation beyond the
coil-stretch transition, one can not neglect the thermal force $\zeta_i$ and
the dynamics of the polymer direction becomes more complicated. We will study
only the statistics of the polymer size $R$ in this case. If the average
polymer size is much smaller than it's nonlinear length, one can assume
the relaxation $\gamma(R)$ to be constant, in which case the equation
(\ref{main}) becomes constant and can be easily analyzed.

Next, it is important to discuss how the chaotic velocity component
is modeled. The statistical properties of the velocity field, observed
in the chaotic turbulence are not known neither from the experimental nor
theoretical points of view. In this paper we will study the simplest model
when the velocity field consists of the strong stationary shear component
and of the weak short correlated chaotic component.
Under this assumptions the velocity gradient matrix has the following form:
\begin{eqnarray}
 \nabla_j v_i = s \delta_{iX}\delta_{jY} + \sigma_{ij} \\
 \langle \sigma_{ij}(t)\sigma_{kl}(t')\rangle= D
 \delta(t-t')(4\delta_{ik}\delta_{jl} -\delta_{il}\delta_{kj}-\delta_{ij}\delta_{kl})
 \label{sigmacorr}
\end{eqnarray}
Here we assume that the shear flow occurs in the $XY$ plane (note that this plane
is stationary in the rotating frame, associated with the polymer). The exact
form of the correlation function (\ref{sigmacorr}) assumes the isotropy of
velocity component, however as we will see this assumption is not important, because
in the case of strong shear component $s\gg D$ the polymer spends most of the
time stretched in the $X$ directions, and it's angular dynamics is determined
only by the $Y$ component of the chaotic velocity field.

In order to simplify the equations on the polymer direction we
parameterize the vector ${\bf n}$ as shown in the Fig.
\ref{fig:orient}. The Eq. (\ref{direct}) acquires the following form
\begin{eqnarray} &&
 \partial_t{\phi} = - s\sin^2\phi + \xi_\phi \,,
 \label{phieq} \\ &&
 \partial_t{\theta} =- s
 \frac{\sin(2\phi)}{2}\sin\theta + \xi_\theta \,,
\label{thetaeq} \end{eqnarray}
where $\xi_\phi$ and $\xi_\theta$ are zero mean random variables related to chaotic
components of the velocity gradient. The statistics of both $\xi_\phi$ and
$\xi_\theta$ can be obtained from the correlation function (\ref{sigmacorr}):
\begin{eqnarray}
 \langle\xi_\theta(t)\xi_\theta(t')\rangle = 4 D \delta(t-t') \\
 \langle\xi_\phi(t)\xi_\phi(t')\rangle = \frac{4 D}{\cos^2\theta} \delta(t-t')
\label{phicorr}
\end{eqnarray}

\section{Statistics of polymer direction}
\subsection{$\phi$-angle distribution}
It follows from (\ref{direct}) that for the stretched polymers the angular dynamics
is decoupled from the dynamics of the polymer size. In the case of $s\gg D$
the regular component will play the main role in the evolution of the polymer
direction. For vanishing chaotic component
$D=0$ the deterministic polymer dynamics can be easily analyzed: there are two
semistable equilibrium states $\phi_{1,2}=0,\pi;\theta_{1,2}=0$, and the polymer
direction vector ${\bf n}$ asymptotically approaches one of these points depending
on it's initial orientation. However when the angle between polymer and equilibrium
direction becomes sufficiently small, the chaotic components $\xi_k$ can not
be neglected, and the polymer dynamics becomes stochastic. After some time the
chaotic component transfers the polymer into instable region, and regular
velocity quickly (on times of order $s^{-1}$) transfers it to the opposite
equilibrium direction. Due to stochastic nature of the chaotic velocity
component one will observe random a-periodic tumbling of polymer. This
phenomena was qualitatively analyzed in \cite{04CKLT} for general velocity
statistics. In this work we focus on the situation where the chaotic flow
is short correlated, such that it's characteristic correlation time $\tau_v$ is
small compared to the time scale, associated with the tumbling effect which as
we will show can be estimated as $\tau_t = (D s^2)^{-1/3} \gg s^{-1}$. This
assumption will allow us to obtain some rigorous results on the stationary and
dynamical statistics of polymer.

When the chaotic component is weak enough $D\ll s$ the polymer spends most of the
time in the stochastic regime, close to the equilibrium point, so that it's
angles are small $\theta,\phi\ll 1$ (we will analyze only one
equilibrium point, because of the symmetry ${\bf n}\to -{\bf n}$).
In this case one can set $\theta=0$ in the correlation function (\ref{phicorr}),
so the dynamics of angle $\phi$ becomes decoupled from everything else, and one can
write the corresponding Fokker-Planck equation:
\begin{equation}\label{fokker}
\left[\partial_t - s\partial_\phi\sin^2\phi - 2D
\partial_\phi^2\right]P(t,\phi) =0
\end{equation}
One of the important questions that should be discussed here are the boundary
conditions, that should be used with this equation. The equation is invariant
under the transformations $\phi \to \phi \pm \pi$. Therefore it would be natural
to use periodic boundary conditions $P(t,-\pi/2)=P(t,\pi/2)$. In this case
there will exist an asymptotic stationary solution of (\ref{fokker})
$P_{st}(\phi)$. Obviously, all angles differing to the value divisible by
$\pi$ are identical to each other in this solution. Another possibility is
to use traditional boundary conditions ${\cal P}(t,\infty)={\cal P}(t,-\infty)=0$.
In this case the angles $\phi$ and $\phi+\pi k$ are not equivalent, and
one can study the statistics of the number of polymer rotations. The main
disadvantage of such approach is that there is no stationary solution of the
Fokker-Planck equation, because the PDF is widening and drifting constantly.
However both of the approaches lead to the same physical results, the following
identity holds between two different PDFs:
\begin{equation}
 P(t,\phi) = \sum_k {\cal P}(t,\phi+\pi k)
\end{equation}
In order to find the stationary PDF $P_{st}(\phi)$ we rewrite the
Fokker-Planck equation in the following form:
\begin{eqnarray}
 \partial_\phi U^{-1}(\phi)\partial_{\phi} U(\phi) P_{st}(\phi)
 =0\\
 U(\phi) = \exp\left[\frac{s}{4D}\phi-\frac{s}{8D}\sin2\phi\right]
\end{eqnarray}
Trivial integration leads to the following expression for PDF:
\begin{equation}
 P_{st}(\phi) =
 \frac{\omega}{D}\int_0^\pi\mathrm{d}\phi
\exp\left[-\frac{s}{4D}\left(\phi-\sin\phi\cos(\phi-2\phi)\right)\right],
\end{equation}
where $\omega$ is the average rotation frequency of the polymer, which is
determined from the normalization condition $\int_0^\pi
P_{st}(\phi)\mathrm{d}\phi=1$ and is given by
\begin{equation}
 \omega = \frac{D\exp\left(\frac{\pi s}{8
 D}\right)}{\pi^2 I_{ix}(x)I_{-ix}(x)},
\end{equation}
where $x=s/8D$. In the most interesting case $s \gg D$, the PDF will
be localized at small angles $\phi\sim (D/s)^{1/3}\ll 1$ and all
expressions are significantly simplified:
\begin{eqnarray}
 \omega = \frac{(D s^2)^{1/3}}{4\cdot3^{1/6}\Gamma(7/6)\sqrt{\pi}} \\
 P_{st}(\phi) =
\frac{\omega}{D}\int_0^\infty\mathrm{d}\phi
\exp\left[-\frac{s}{8D}\phi(\phi-2\phi)^2-\frac{s\phi^3}{24 D}\right]
\end{eqnarray}
One can see, that the PDF is asymmetric and has wide algebraic tails:
\begin{eqnarray}
 \langle\phi\rangle =
 \left(\frac{D}{s}\right)^{1/3}\frac{\sqrt{\pi} 3^{1/3}}{\Gamma(1/6)} \\
 P_{st}(\phi) \sim \frac{1}{16
\cdot3^{1/6}\Gamma(7/6)\sqrt{\pi}}\left(\frac{s}{D}\right)^{2/3}\frac{1}{\phi^2},
\end{eqnarray}
where the last asymptotic is valid in the intermediate region $(D/s)^{1/3}\ll|\phi|\ll 1$
The asymptotic $P_{st}\propto \phi^{-2}$ corresponds to the non-zero
probability flux at $|\phi|\to\infty$ which is a manifestation of the
described above tumbling effect. Positive value of the average angle
$\phi$ shows, that the polymer spends most of the time in the region
$\phi > 0$ in agreement with the qualitative analysis presented in the
first sections of this paper. One can see, that the polymer spends most of the
time in the region of small angles, therefore the only relevant velocity
component $v_z$. The assumption of isotropic statistics of chaotic
velocity component is not therefore significant for the qualitative results of
this paper.

\subsection{Tumbling time statistics}
\label{timestat}
In this section we will calculate the probability distribution function of the
time intervals between consequent tumblings. Such PDF can be directly measured
experimentally. For this problem it is naturally to use the non-stationary PDF
${\cal P}(t,\phi)$. We will define the tumbling process by a polymer direction
``trajectory'' starting at $\phi=\pi/2$ and reaching $\phi=-\pi/2$ at time
$T$. In this case the probability of finding the polymer inside this region is
given by
\begin{equation}
 p(t) = \int_{-\pi/2}^{\pi/2} {\cal P}(t,\phi) \mathrm{d}\phi.
\end{equation}
Where the initial condition is ${\cal P}(t,\phi)=\delta(\phi-\pi/2+0)$,
so $p(0)=1$. We substitute ${\cal P}(t,\phi) = U^{1/2}(\phi)
\Psi(t,\phi)$. In this case the evolution of $\Psi$ is determined by
the one-dimensional Schroedinger equation in imaginary time:
\begin{eqnarray}
 \partial_t \Psi = - \hat{H}\Psi \\
 \hat{H} = -2D \partial_\phi^2 + \frac{s^2}{8D}\sin^4\phi-s \sin\phi\cos\phi
\end{eqnarray}
Now it is possible to use the quantum-mechanical analogy. The hamiltonian
$\hat{H}$ formally describes a particle in periodic potential with the period
$\pi$. General solution of this problem looks like
\begin{equation}
 \Psi(t,\phi) = \sum_n\int\mathrm{d}p
 \Psi_{np}(\phi)\Psi_{np}^*(\pi/2)\exp(-E_n(p)t),
\end{equation}
where $p$ is particle's quasimomentum, and index $n$ enumerates the Brillouin
zone number. In this potential the classical minimums are separated by large
barriers. And for $s\gg D$ one can use the tight-binding
method \cite{abrikosov}. The following approximate relations hold for the
spectrum:
\begin{eqnarray}
 E_n(p) = \epsilon_n - \nu\cos(\pi p) \\
 \Psi_{np}(\phi) = \sum_k \exp(i\pi k p) \psi_n(\phi-k\pi),
\end{eqnarray}
where $\psi_n,\epsilon_n$ is the spectrum, which is formed near
classical minimums, when the tunneling processes are neglected.
$\nu$ is exponentially small band width. Therefore at large times
the main asymptotic of $p(t)$ will be determined by the ground state
energy $\epsilon_0$:
\begin{equation}
 p(t) \propto \exp(-\epsilon_0 t),\qquad t\to\infty
\end{equation}
One can easily check that this energy is given by $\epsilon_0 = c
(Ds^2)^{1/3}$, where $c$ is a constant of order unity. Indeed, the classical
minima is situated in the region of small angles $|\phi|\ll 1$, so one can
use Taylor expansions of trigonometric functions. After the substitution
$\phi = (D/s)^{1/3}$ one obtains the Hamiltonian
\begin{equation}
 \hat{H} = (D s^2)^{1/3}\left[-2 \partial_x^2+\frac{x^4}{8}-x\right].
\end{equation}
The operator in square brackets contains no dimensionless parameters, and
therefore it's eigenvalues will be of order unity. The body of the PDF is
also situated in the region of tumbling periods of order $T\sim (Ds^2)^{-1/3}$.
Left tail of tumbling time PDF $T\ll (Ds^2)^{-1/3}$ is determined by rare
trajectories, which turn the polymer by angle $\pi$ at small times $T$. In
order to find the optimal form of such trajectories we will functional
integral representation of the transition probability
\begin{equation}
 p(T) \propto \int {\cal D}\phi \exp\left[-\frac{1}{8 D}\int\mathrm{d}t
 (\dot{\phi}+s\sin^2\phi)^2\right]
\end{equation}
The integration is taken over trajectories with boundary conditions
$\phi(0)=\pi/2,\phi(T)=-\pi/2$. For small $T \ll (Ds^2)^{-1/3}$ the
probability is determined by the action $A$ on the optimal trajectory
$p(T)\sim\exp(-A)$ with exponential accuracy. Variation of the effective action
leads to the following equation on instanton:
\begin{equation}
 \ddot{\phi}= s^2 \sin^3\phi\cos\phi
\end{equation}
This mechanical problem can be easily solved, and one obtains the following
relation between the tumbling time and effective particle energy:
\begin{widetext}
\begin{equation}
 T = \int_{-\pi/2}^{\pi/2}\frac{\mathrm{d}\phi}{\sqrt{2 E + s^2\sin^4\phi}}
 = \left[\frac{8}{E(2E+s^2)}\right]^{1/4}
 K\left(\frac{1}{2}-\sqrt{\frac{E}{4E+2s^2}}\right),
\end{equation}
where $K(x)$ is elliptical integral of the first kind. The action in this case
is given by
\begin{equation}
 A =  \frac{E T}{4D}
 +\frac{s^2}{4D}\int\mathrm{d}\phi
 \frac{\sin^4\phi}{\sqrt{2E+s^2\sin^4\phi}}
 = \frac{E T}{4D}
 +\frac{3\pi s^2}{32D\sqrt{2E}}
  \,_3F_2\left(\frac{1}{2},\frac{5}{4},\frac{7}{4};\frac{3}{2},2;-\frac{s^2}{2 E}\right)
\end{equation}
\end{widetext}
Here we omitted the constant term $2 s \int \dot{\phi}\sin^2\phi
\mathrm{d}t = \pi s/(8 D)$ because it is cancelled by the normalization
constant. Because of additional time scale $s^{-1}$ there are two different
asymptotes of $p(t)$. In the case of $s T \ll 1$ one has $s \ll \sqrt{2 E}$
and $E = \pi^2/(2T^2)$. In this case
\begin{equation} \label{smallest}
 A = \frac{\pi^2}{8 D T}
\end{equation}
In another limiting case $s^{-1}\ll T \ll (Ds^2)^{-1/3}$ the energy is given by
$E = 8 K^4(1/2)/(s^2 T^4)$ and the action has the following form:
\begin{equation} \label{inter}
  A = \frac{2 K^4(1/2)}{3 D s^2 T^3}
\end{equation}
The intermediate asymptotic (\ref{inter}) is a function of the product
$D s^2 T^3$ because it is determined by the dynamics in the region of small
angles. Therefore it will be universal, in the sense mentioned above. In
contrary, the asymptotic at the smallest times (\ref{smallest}) does not depend
on $s$ at all, because such small times can be reached only due to very rare
fluctuation of chaotic velocity field. Therefore, this asymptotic strongly
depends on the assumption of isotropic velocity statistics, and is not
universal.

\subsection{$\theta$-angle distribution} \label{thetasec}
As it was shown in \cite{04CKLT} there are two contributions to the
intermediate right tail of the $\theta$-angle distribution
$(D/s)^{1/3}\ll\theta\ll 1$. The first one is coming from the region
of deterministic regions where $\phi\sim 1$, where the angular
dynamics is determined by the regular terms. The algebraic tail in
this case is proportional to $\theta^{-2}$, however there is also a
non-universal algebraic part, which comes from the stochastic region
of $\phi \sim (D/s)^{1/3}$ and is determined by the statistical
properties of the random velocity field. In this section we will be
interested in this part, and will obtain the connection between the
exponent and the entropy function of the random velocity process. In
the region $\theta\ll 1$ the equation (\ref{thetaeq}) can be easily
solved:
\begin{equation}
 \theta(t) = \int_0^\infty \mathrm{d}\tau
\exp\left(-\frac{s}{2}\int_{t-\tau}^t\sin2\phi(t')\mathrm{d}t'\right)
\xi_\theta(t-\tau)
\end{equation}
As we already know the random process $\phi(t)$ is stationary and
independent from $\xi_\theta(t)$. In this case the expression for $\theta$
can be rewritten in the form
\begin{eqnarray}
 \theta = \int_0^\infty \mathrm{d}\tau \exp(-\varrho(\tau))\xi_\theta(\tau)\\
 \varrho(\tau) = \frac{s}{2}\int_0^\tau\sin2\phi(t)\mathrm{d}t
\end{eqnarray}
In order to obtain the PDF $P(\theta)$ we first average over the noise
$\xi_\theta$:
\begin{eqnarray} \label{thetadist}
 P(\theta|\varrho) = (2\pi A )^{-1/2} \exp\left[-\frac{\theta^2}{2 A}\right] \\
 A = 4D \int_0^\infty \mathrm{d}\tau\exp[-2\varrho(\tau)].
\end{eqnarray}
Here $P(\theta|\varrho)$ is the PDF of $\theta$ for a fixed realization of the
process $\varrho(t)$. Due to positive value of $\langle
\dot{\varrho}\rangle\sim (Ds^2)^{1/3}$ the body of $P(\theta)$ is situated in
the region of small angles $\theta\sim (D/s)^{1/3}\ll 1$. Tails of the PDF are
determined by large deviations of negative $\varrho(t)$. Assuming that the
process $\varrho(t)$ reaches it's most negative value to the time $\tau^*$, such
that $\varrho(\tau^*)=-\varrho^*$ and $\rho^*\gg 1$ one can estimate the value
of $A$ with exponential accuracy as $A\sim (D/s)^{1/3}\exp(2\varrho^*)$. The
characteristic correlation time of $\varrho(t)$ is $\tau_c = (Ds^2)^{-1/3}$,
so for large $\tau^*\gg \tau_c$ one can use the results of large deviations
theory \cite{85Ell} which predict the following scaling for the tails of
$\rho^*$ PDF:
\begin{equation}
 P(\varrho^*|\tau^*) = \exp\left[-\frac{\tau^*}{\tau_c}
 S(\frac{\varrho^*\tau_c}{\tau^*})\right],
\end{equation}
where $S(x)$ is the entropy function, which form can not be found analytically
in general case. One can now find the most probable time $\tau^*$ by maximizing
the above probability over $\tau^*$. This leads one to the expression
$\tau^* = \tau_c \varrho^*/x^*$, where $x$ is found from the following
equation:
\begin{equation}
 S(x^*) = x^* S'(x^*)
\end{equation}
The entropy function is of order unity, so one can expect the same from $x^*$.
The asymptotic of $\varrho^*$ PDF is therefore given by
\begin{equation}
 P(\varrho^*) \sim \exp(-\varrho^* S'(x^*))
\end{equation}
and after averaging (\ref{thetadist}) over $\varrho^*$ one obtains
the following asymptotic of $\theta$ PDF:
\begin{equation}
 P(\theta) \propto |\theta|^{-S'(x^*)},\qquad (D/s)^{1/3}\ll|\theta|\ll 1
\end{equation}
One can see that the tails are algebraic as in the case of the angle $\phi$,
however the exponent is now non-universal, and depends on the statistical
properties of the velocity field.

\section{Statistics of polymer elongation}
The tails of the polymer size PDF can be studied in the similar way as in
section \ref{thetasec}. The explicit solution of the dynamical equation
\ref{main} in the case of linear relaxation force has the following form
\begin{eqnarray}
 R_i(t) = \int_0^\infty \mathrm{d} t \exp\left[-\gamma (t-t')\right]
 W_{ij}(t,t') \zeta_j(t') \\
 W = T\exp\left[\int_{t'}^t \tilde{\sigma}(\tau)\mathrm{d}\tau\right]
\end{eqnarray}
where $\tilde{\sigma}_{ij} = \nabla_j v_i$ is the velocity gradient matrix.
In order to obtain polymer size PDF we first average over the thermal
Langevin force $\xi_i(t)$:
\begin{eqnarray}
 P({\bf R}|\tilde{\sigma}) \propto \exp\left[-\frac{1}{2}{\bf R}^T I^{-1}{\bf R}\right] \\
 I = \kappa\int_0^\infty \mathrm{d} t' W^T(t')W(t)\exp(-2 \gamma t)
\end{eqnarray}
where $W(t)= W(t,0)$, and $P({\bf R}|\tilde{\sigma})$ stands for the PDF with a
fixed realization of the process $\tilde{\sigma}(t)$. At large enough times
$t\gg \tau_c$ the eigenvalues of the matrix $W^T W$ become widely separated and
the absolute value of the end-to-end vector ${\bf R}$ is determined by the largest
eigenvalue $I_1$:
\begin{eqnarray}
 P(R|\tilde{\sigma}) \propto \exp\left[-\frac{R^2}{2 I_1}\right]
\end{eqnarray}
It can be easily shown (see e.g. \cite{99BF}) that for large times, when the
eigenvalues $\lambda_{i}$ of $W^T W$ are widely separated
$\lambda_1\gg\lambda_2\gg\lambda_3$ the dynamics of the largest one
$\lambda_1=\exp(2\rho)$ is described by the following equation:
\begin{eqnarray} \label{rhoeqq}
 \dot{\rho} = \frac{s}{2} \cos^2\theta\sin2\phi+ 6D+\xi_\rho, \\
 \langle\xi_\rho(t)\xi_\rho(t')\rangle = 2 D \delta(t-t') \\
 \label{rhoeq}
\end{eqnarray}
where $\xi_\rho$ is obtained from the chaotic velocity correlation function
(\ref{sigmacorr}).  The eigenvalue $I_1$ is then
given by the expression
\begin{eqnarray}\label{I1}
 I_1 = \kappa \int_0^\infty \mathrm{d}t\exp\left[2\rho(t)-2\gamma t\right]
\end{eqnarray}
Like in the previous section $\rho(t)$ is an integral of stationary random
process with the correlation time of order $\tau_c = (Ds^2)^{-1/3}$, and large
deviations of $I_1$ are determined by the large deviations of $\rho(t)$. Assuming
that the integral (\ref{I1}) is determined by one saddle point $\tau^*$
and  can be estimated like $I_1 \propto\exp(2\rho^*-2\gamma\tau^*)$ where
$\rho^* = \rho(\tau^*) $. The asymptotic of $\rho^*$ PDF for a fixed value of
$\tau^*$ has the following form:
\begin{equation}
 P(\rho^*|\tau^*) \propto \exp\left[-\frac{\tau^*}{\tau_c}
 S_\rho\left(\frac{\rho^*\tau_c}{\tau^*}\right)\right]
\end{equation}
One can now find the optimal value of $\tau^* = \tau_c \rho^*/x$. The
coefficient $x$ is found from the following equation:
\begin{equation}
 S_\rho(x) -x S_\rho'(x) + \gamma \tau_c S_\rho'(x)
\end{equation}
As long as we study the linear region beneath the coil-stretch transition
we have $\gamma \tau_c > 1$. The tail of PDF will be algebraic like in case of
$\theta$ angle: $P(R) \propto R^{-1-\alpha}$, and the value of $\alpha$ can be
determined for large values of $\gamma\tau_c \gg 1$. Large deviations of $R$ are
determined by the region where thermal Langevin forces can be neglected and one
can use the equation (\ref{rhoeqq}). We are interested in the asymptotic behaviour of
the polymer size moments $M_q(t) = \langle R^q(t)\rangle\propto \exp(A_q t)$. In
this case the value of $\alpha$ will be determined from the equation $A_\alpha =0$.
Integrating out the function $\xi_\rho$ one can rewrite $M_q$ in the following
form:
\begin{eqnarray}
 M_q = \exp\left[D q^2 t - \gamma q t\right] \int\mathrm{d}\phi
 Z_q(\phi,t) \\
 Z_q = \left\langle \exp[q\varrho(t)]\delta(\phi-\phi(t))\right\rangle
\end{eqnarray}
where the angle brackets here stand for the averaging over the process
$\phi(t)$. The function $Z_q$ obeys the equation
\begin{equation}
 \partial_t Z_q = \left[2D \partial_\phi^2 + s
 \partial_\phi\sin^2\phi+\frac{qs}{2}\sin2\phi\right]Z_q
\end{equation}
The only difference from the Fokker-Planck equation \ref{fokker} only in the
last term. We will follow the same procedure as in the section \ref{timestat}.
Substituting $Z_q = U(\phi)\Psi(t,\phi)$ we obtain the imaginary time
Schroedinger equation
\begin{eqnarray}
 \partial_t \Psi = - \hat{H_q}\Psi \\
 \hat{H_q} = -2D \partial_\phi^2 + \frac{s^2}{8D}\sin^4\phi+(q-1)s \sin\phi\cos\phi
\end{eqnarray}
This equation can not be solved in the case $q\sim 1$, however for $q\gg 1$
the solution can be easily found. In this case the main exponential asymptotic
at large times is $Z_q(t) = \propto \exp(-\epsilon(q)t)$, where $\epsilon(q)$ is
the ground state energy. For $q\gg 1$ the main contribution to $\epsilon(q)$
will be equal to the value of classical minima of the potential. After some
simple algebra we obtain
\begin{eqnarray}
& \epsilon(q) \sim -3\cdot 2^{-5/3} \left(q^4 Ds^2\right)^{1/3} ,\qquad 1 \ll
q\ll s/D\\
& \epsilon(q) \sim qs/2,\qquad q\gg s/D
\end{eqnarray}
Finally we have
\begin{equation}
 A(q) = Dq^2-\gamma q -\epsilon(q)
\end{equation}
and the critical value $\alpha$ will depend on the dimensionless parameter
$\gamma/s$:
\begin{eqnarray}\label{res}
 \alpha = \frac{81}{32}\frac{\gamma^3}{D s^2},\qquad \gamma \ll s\\
 \alpha = \frac{\gamma}{D},\qquad \gamma\gg s \label{last}
\end{eqnarray}
The last expression (\ref{last}) coincides with the value of exponent for pure
isotropic chaotic velocity, because in the case $\gamma \gg s$ large polymer
size fluctuations are determined by the rare fluctuations of the chaotic
component, when the flow has a strong elongation component with the lyapunov
exponent $\lambda > \gamma$ for a long time. However the result (\ref{res})
shows that in the case $\gamma \ll s$ the shear component can significantly
broaden the tails of the polymer size PDF. This fact is nontrivial because the
regular shear component itself can not lead to the exponential polymer
elongation, and non-trivial exponent comes from the combined effect of chaotic
and regular component.

\section{Conclusions}
In conclusion we would like to repeat the main results of this paper. A polymer
molecule injected in the chaotic flow with the strong mean shear component
becomes strongly stretched in the shear direction and experiences the a-periodic
tumbling in the plane of the shear flow. It is shown, that stationary distribution
of the polymer orientation angles has intermediate algebraic asymptotics. The tails
of the $\phi$-angle distribution behave like $\phi^{-2}$ and correspond to the
constant flux solution of the Fokker-Planck equation. Existence of these
asymptotics has been reported in previous papers regarding the dynamics of
molecules and small particles in the shear flows \cite{book,HinchLeal}. However
the algebraic asymptotic of the $\theta$-angle distribution has not
been reported in any previous papers to our knowledge. We would like to stress,
that in contrary to the PDF of the $\phi$ angle, the asymptotic behaviour of the
$\theta$-angle distribution is not universal and depends on the statistical
properties of the chaotic velocity component. Therefore in our opinion
measurement of the scaling exponent of the $\theta$-angle distribution is a
perfect tool for the investigation of the statistical properties of the chaotic
velocity. Another quantity which can be measured by direct polymer observations
is the PDF of the time periods between consequent polymer tumblings. The
equilibrium direction of the polymer, which is oriented along the shear direction
is semistable. This property leads to frequent polymer tumblings, so that the
characteristic time between tumblings is of order $\tau \sim \tau_c = (Ds^2)^{-1/3}$.
The situation would be different in the case of vesicles or non-spherical particles
in the shear flow \cite{HinchLeal}, where the tumbling rate would be exponentially
small. In the fourth part of this paper we make the prediction on the polymer size
distribution. We show, that the existence of the strong shear component leads
to the significant broadening of the size distribution compared to the isotropic
case, which was considered in \cite{00BFL}. This effect is to our opinion rather
non-trivial, because the shear component itself can not lead to the exponential
elongation of the polymer and the distribution broadening is the combined effect
of the chaotic and regular velocity components. Finally, we would like to mention
that all results of this paper were obtained under the assumption of isotropic and
short-correlated chaotic velocity flow. While the first assumption, as it was
discussed throughout the paper is irrelevant for most of the results, the finite
correlation time of the velocity flow can lead to the change of some quantitative
results of this paper. However we believe (see \cite{04CKLT} for the more detailed
discussion) that most qualitative predictions will remain valid if the velocity
correlation is smaller or comparable to the characteristic time scale of the polymer
dynamics $\tau_c$.

Author would like to thank M. Chertkov, I. Kolokolov and V. Lebedev
for numerous of inspiring discussions. This work was supported by
the Dynasty foundation and RFBR grant 04-02-16520a.

\end{document}